\documentclass[11pt]{article}
\usepackage{moriond,epsfig}

\def\Journal#1#2#3#4{{#1} {\bf #2}, #3 (#4)}

\def\NPA{{\em Nucl. Phys.} A}
\def\NPB{{\em Nucl. Phys.} B}
\def\PLB{{\em Phys. Lett.}  B}
\def\PRL{\em Phys. Rev. Lett.}
\def\PRD{{\em Phys. Rev.} D}

\def\EPJC{{\em Eur. Phys. J.} C}
\def\PR{\em Phys. Rep.}

\def\be{\begin{equation}}
\def\ee{\end{equation}}
\def\bea{\begin{eqnarray}}
\def\eea{\end{eqnarray}}

\begin{document}
\vspace*{4cm}
\title{SUPPRESSION OF FORWARD PION CORRELATIONS IN d+Au INTERACTIONS AT STAR}

\author{ E. BRAIDOT, FOR THE STAR COLLABORATION}

\address{Institute for Subatomic Physics, Utrecht University, Princetonplein 5,\\
3584 CC Utrecht, The Netherlands}

\maketitle\abstracts{
During the 2008 run RHIC provided high luminosity in both p+p and 
 d+Au collisions at $\sqrt{s_{NN}}=200GeV$. Electromagnetic calorimeter
 acceptance in STAR was enhanced by the new Forward Meson 
 Spectrometer (FMS), and is now almost contiguous from $-1<\eta<4$ 
 over the full azimuth. This large acceptance provides sensitivity to 
 the gluon density in the nucleus down to $x\approx 10^{-3}$. 
 Measurements of the azimuthal correlation between a forward $\pi^0$ 
 and an associated particle at large rapidity are sensitive to the 
 low-$x$ gluon density. Data exhibit the qualitative features expected
 from gluon saturation. A comparison to calculations using 
 the Color Glass Condensate (CGC) model is presented.}

\section{Introduction}
Forward acceptance at STAR has been drastically improved in the last few years with the development and commissioning of the Forward Meson Spectrometer (FMS). The FMS is a high granularity electromagnetic calorimeter whose purpose is to measure photons from decays of forward neutral mesons. It extends the fully azimuthal, electromagnetic capability of STAR into the pseudorapidity region $2.5<\eta<4.0$, making the overall coverage almost hermetic in the wide $-1.0<\eta<4.0$ interval. This allows measurements of correlations of different species of forward- and mid-rapidity particles, over a broad $\Delta\eta\times\Delta\varphi$ range. 

One of the main purposes of the FMS \cite{future} is to characterize the $p_{T}$ dependence of the di-pion azimuthal correlations as a means to search for the onset of saturation effects \cite{satu} in the nuclear gluon distribution. It is believed that, as the energy grows, non-linear effects must be included in the nuclear wave-function in order to tame the otherwise divergent rise of the gluon density. At very low values of the longitudinal momentum fraction $x$ of the probed gluon, as accessed in high-energy collisions, the occupation numbers become large, allowing gluon recombination that eventually leads to saturation.  
Saturation is expected to be revealed at RHIC in d+Au collisions, where large densities of nuclear gluons are probed with a much simpler final state than heavy ion collisions. The FMS, facing the deuteron beam direction, provides measurements of forward hadron production in the rapidity region that selects small-$x$ gluons in the nucleus. It probes $x$ down to $x\approx10^{-3}$ for inclusive particle production at $\langle\eta\rangle\approx3$ at $\sqrt{s_{NN}}=\textrm{200 GeV}$ \cite{GSV}, well into the range where saturation effects are expected to set in. 

\begin{figure}
\includegraphics[width=0.5\textwidth, height=7cm]{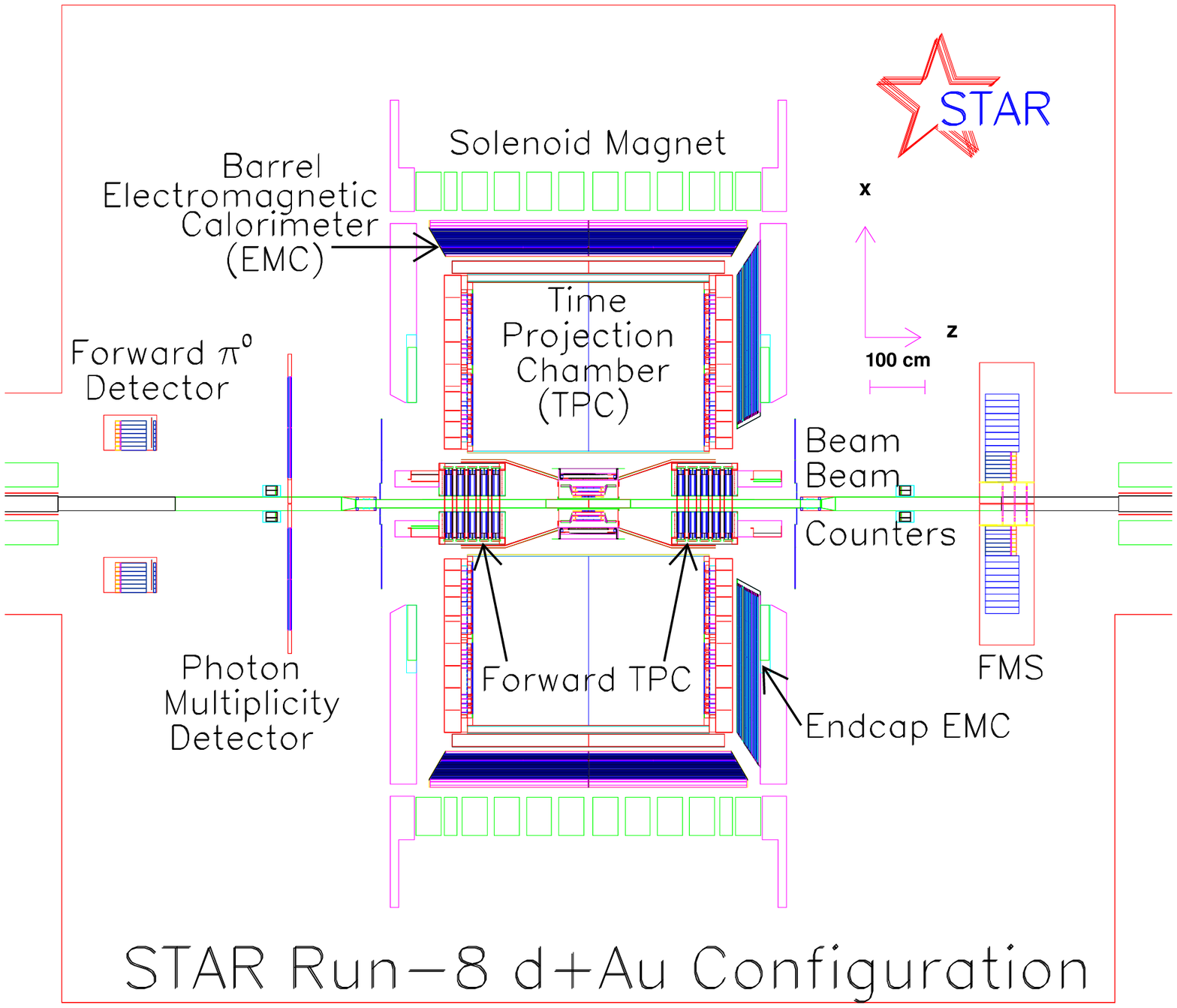}
\includegraphics[width=0.5\textwidth, height=7.5cm]{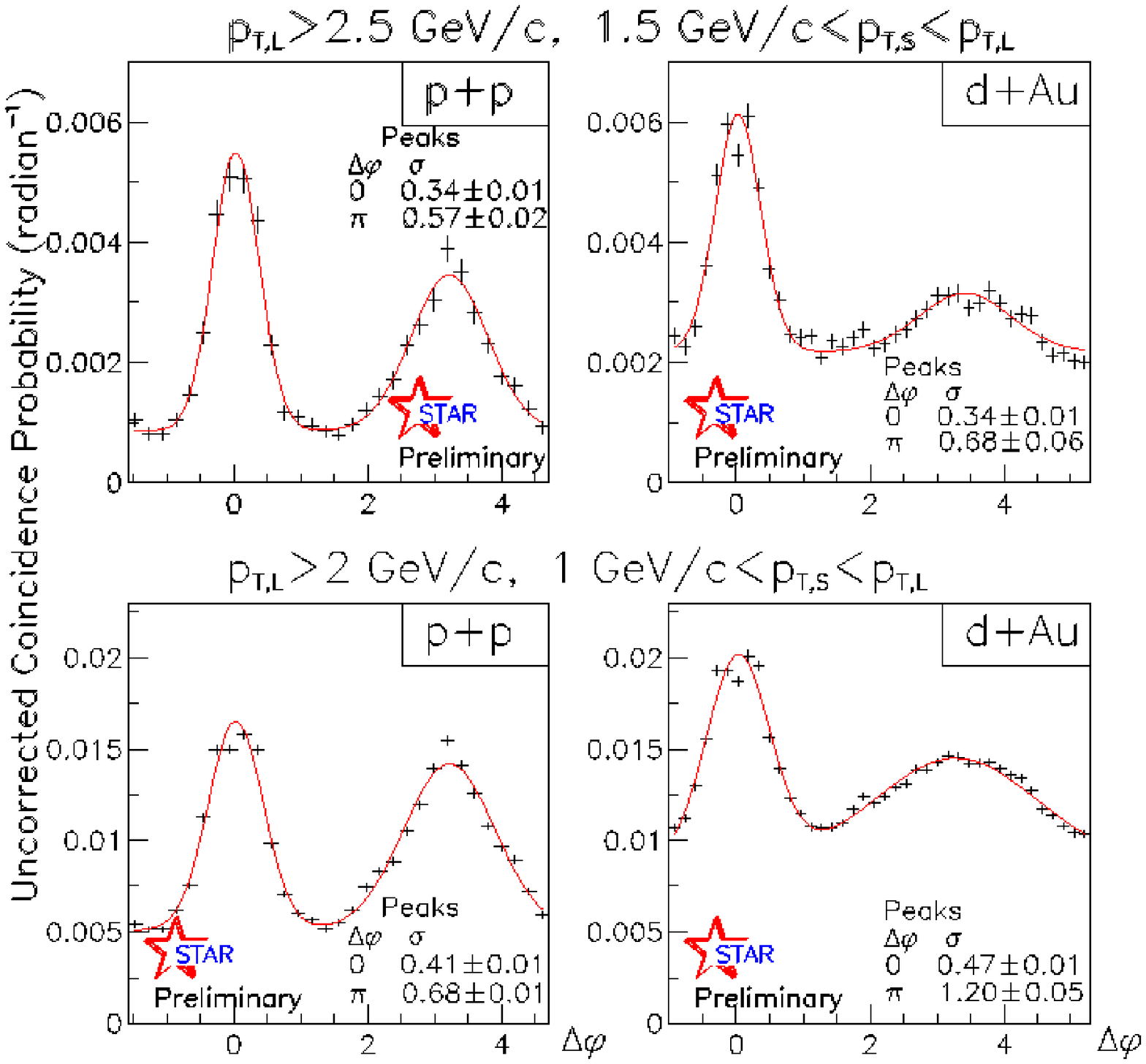}
\caption{On the left: schematic view of the STAR hall. Gold beam is coming from the West (right in figure) side. On the right: uncorrected coincidence probability versus azimuthal angle difference between two forward neutral pions. Different $p_{T}$ cuts are indicated inside the figure. The left (right) column shows p+p (d+Au) data with statistical errors. Data are fit with a constant plus two gaussian functions (in red).}\label{fmsfms}
\end{figure}

\section{Azimuthal correlation analysis}

Models try to describe forward hadron production from dense nuclear targets by including non-linear effects. One approach \cite{Qiu} extends perturbative QCD factorization by adding contributions of quarks scattering coherently off multiple nucleons, leading to an effective shift in the gluon $x$. In the Color Glass Condensate (CGC) model \cite{satu,saturation} the low-$x$ gluon density is saturated and non-linearities are treated classically.
In both cases the $2\rightarrow2$ picture of elastic parton scattering process is replaced by a  $2\rightarrow\textrm{many}$ picture, where the probe scatters coherently off the dense gluon field of the target, which recoils collectively. In the CGC model, the transverse momentum of a jet produced by the large momentum parton (most likely a valence quark) in the deuteron is balanced by many gluons in the nucleus. To quantify this effect, it is instructive to compare the azimuthal angular correlation ($\Delta\varphi$) between the forward $\pi^{0}$ and coincident hadrons. In p+p collisions, one expects to see a peak centered at $\Delta\varphi=\pi$, representing the elastic back-to-back scattering contribution. In forward d+Au collisions, on the other hand, non-linear contributions are expected to cause a loss of correlation between the two particles, leading to a broadening of the back-to-back peak, and eventually to its disappearance (monojet).

A systematic plan of azimuthal correlation measurements has been pursued at STAR using forward prototype calorimeters (Forward Pion Detector, FPD/FPD++) and, more recently, the FMS. The objective has been to probe if the boundaries of the saturation region are accessible at RHIC energies \cite{Albacete} and establish the effect on particle production. Azimuthal correlations between a forward $\pi^{0}$ and a mid-rapidity particle were first measured with the FPD \cite{fwd} and recently confirmed, and extended, with the FMS \cite{braidot}. Comparison of $\Delta\varphi$ between p+p and d+Au shows a significant broadening in the back-to-back peak in d+Au. Such effect appears to be stronger as the transverse momentum of the particles decreases, as expected by saturation models. One can approach the saturation region by lowering the $x$ value of the probed gluon by requiring both leading and associated particles to be detected in the forward region, now possible due to the FMS wide acceptance. Results from this analysis, where the lowest $x$ is probed, are shown.

\section{Results and Systematics}
\def\imagetop#1{\vtop{\null\hbox{#1}}}
\begin{figure}
\begin{tabular}{c c c}
\imagetop{\includegraphics[height=0.3\textwidth]{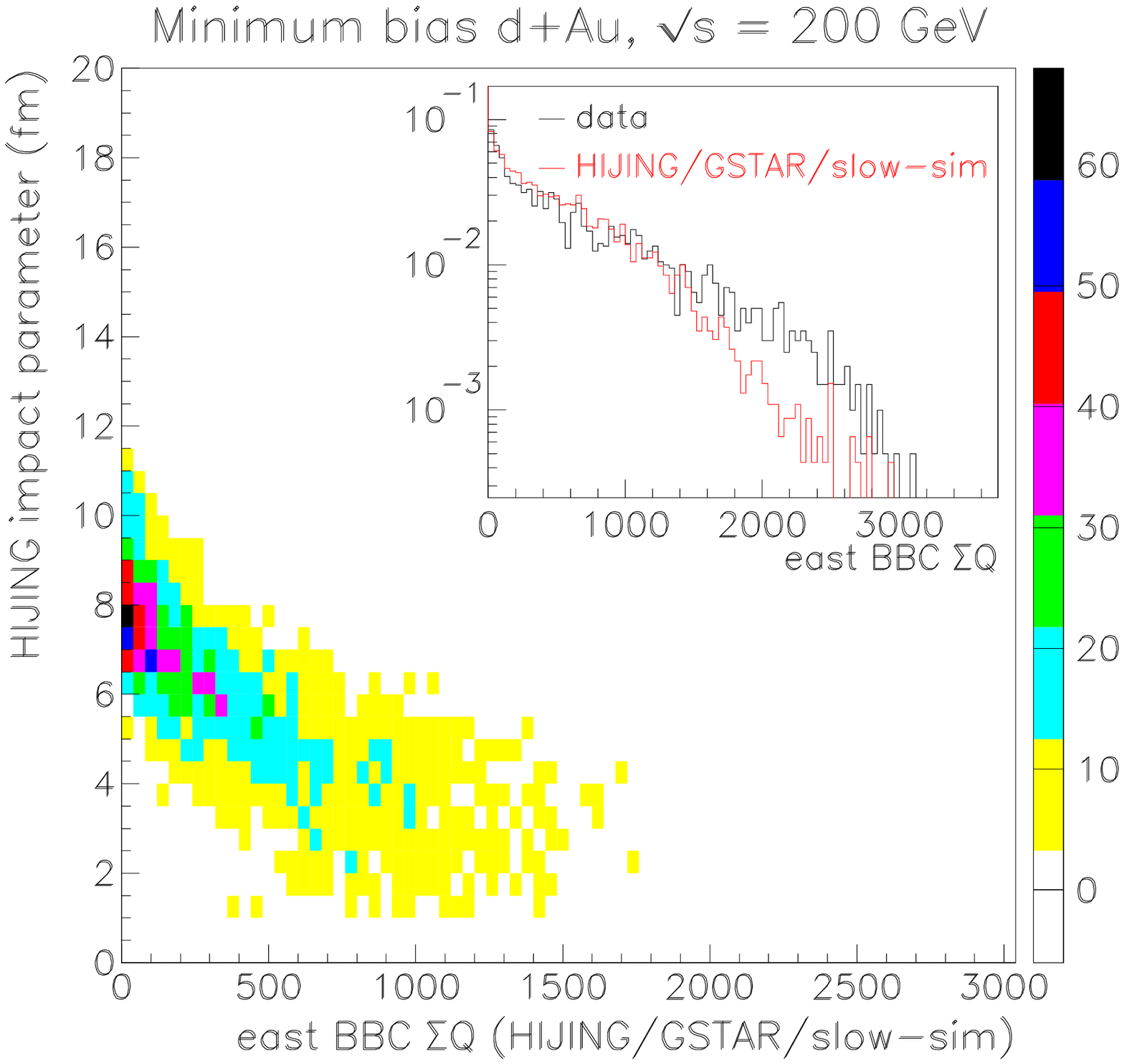}}
  \imagetop{\includegraphics[height=0.3\textwidth]{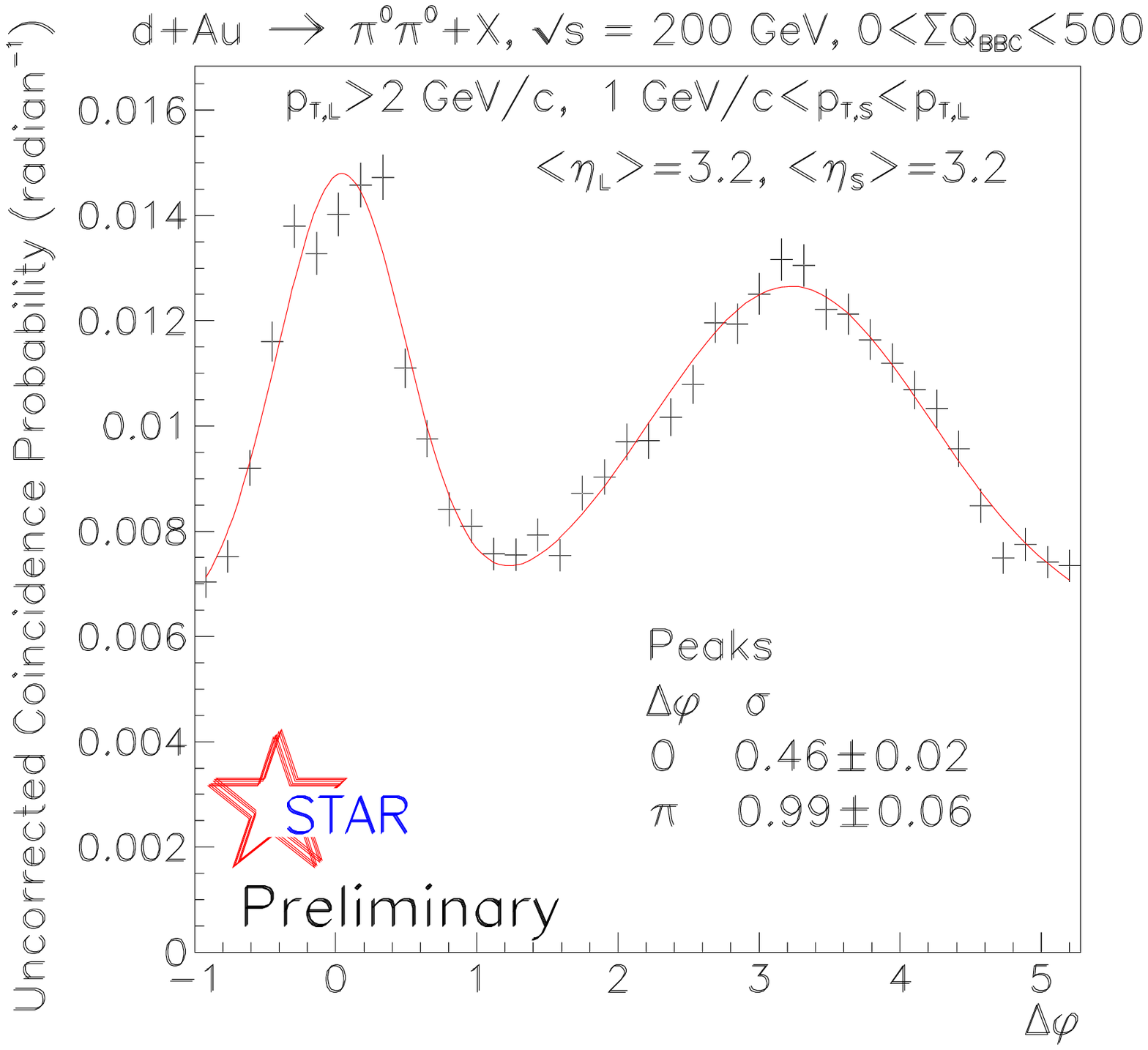}}
\imagetop{\includegraphics[height=0.285\textwidth]{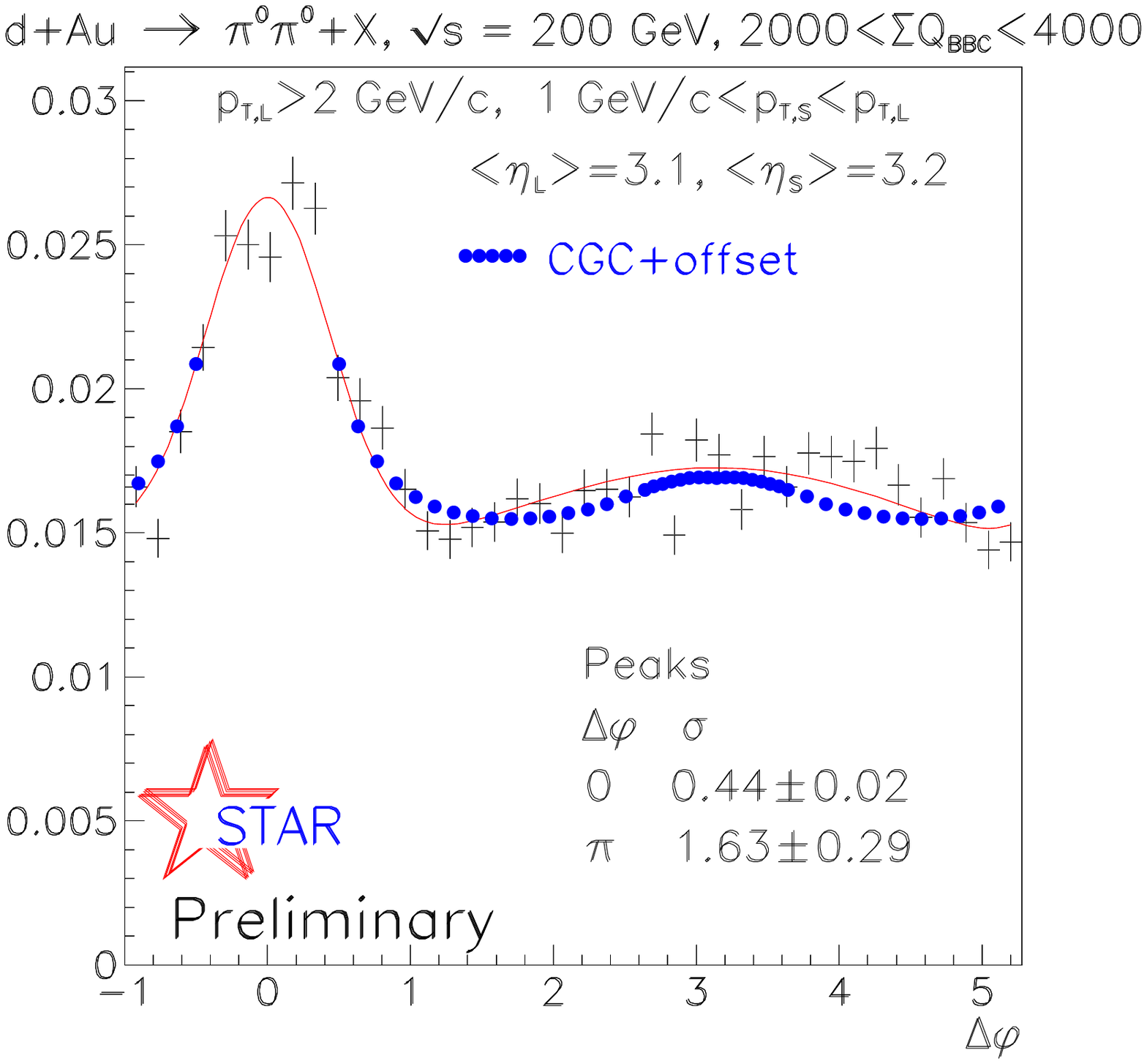}}
\end{tabular}
\caption[]{Left: HIJING impact parameter versus charge sum as recorded by the STAR BBC for simulated minimum bias events. Comparison of charge distribution with data is in the inset. Following: uncorrected coincidence probability versus azimuthal angle difference between two forward neutral pions in peripheral (center) and central d+Au collisions (right). Data are shown with statistical errors and fit with a constant plus two gaussian functions (in red). CGC expectations \cite{cyrille} have  been superimposed (in blue) to data for central d+Au collisions.}
\label{central}
\end{figure}

Centrality averaged azimuthal correlations between two forward $\pi^{0}$ candidates have been measured for different $p_{T}$ selections. Results are shown on the right-hand side of figure \ref{fmsfms}. Neutral pions are reconstructed from pairs of photon clusters, found within the FMS fiducial volume ($2.5<\eta<4.0$), that present an invariant mass in the interval $0.05<M_{\gamma\gamma}<0.25$ GeV/c$^{2}$. The pair with the largest $p_{T}$ is selected as the leading (trigger) $\pi^{0}$ and its azimuthal coordinate is compared inclusively with those of all the other (associated) $\pi^{0}$ candidates. The (efficiency uncorrected) probability to find an associated $\pi^{0}$ per triggered event presents two contributions. 
The peak centered in $\Delta\varphi=0$ (near-side peak) represents the contribution from pairs of neutral pions from the same jet. It is not expected to be affected by gluon saturation, hence providing us with a useful tool to check the effective amount of the broadening in the away-side contribution. This second peak, centered at $\Delta\varphi=\pi$, represents the back-to-back contribution to the coincidence probability and it is expected to disappear in going from p+p to d+Au if saturation sets in. Data are fit with a constant plus two gaussians centered at $\Delta\varphi=0$ and $\Delta\varphi=\pi$ respectively. Figure \ref{fmsfms} shows how the width of the near-side peak is not changing from p+p to d+Au, while the away-side peak presents significative broadening, with an effect larger than that found in forward + mid-rapidity particle correlations \cite{braidot}. The $p_{T}$ dependence of the broadening effect is studied by applying two different selections to the data. A lower $p_{T}$ cut for both trigger ($2.0$ GeV/c$<p_{T}^{(trg)}$) and associated pions ($1.0$ GeV/c$<p_{T}^{(assc)}<p_{T}^{(trg)}$) shows a stronger broadening of the signal width ($\sigma_{dAu}-\sigma_{pp}=0.52\pm0.05$) than a more restrictive cut ($2.5$ GeV/c$<p_{T}^{(trg)}$, $1.5$ GeV/c$<p_{T}^{(assc)}<p_{T}^{(trg)}$, $\sigma_{dAu}-\sigma_{pp}=0.11\pm0.06$), as expected from saturation models.

Broadening effects are expected to be more significant when the more central part of the nucleus is probed. In order to disentangle peripheral from central collisions, the sum of charges ($\sum Q_{BBC}$) has been recorded using the east side of the STAR Beam-Beam Counter (BBC) that faces the Au beam. This provides a measure of the multiplicity of the event, which is correlated with the impact parameter in the collision, as shown on the left-hand panel of figure \ref{central} for minimum bias events, as simulated by HIJING 1.383 \cite{hijing}. The multiplicity selections for peripheral ($0<\sum Q_{BBC}<500$) and central d+Au collisions ($2000<\sum Q_{BBC}<4000$) result in $\langle b\rangle=6.8$ fm and $\langle b\rangle=2.7$ fm respectively. Figure \ref{central} also shows azimuthal correlations for the lower $p_{T}$ selection, using the charge sum to discriminate between peripheral (central panel) and central (right-hand panel) d+Au collision. Peaks on the near-side appear nearly unchanged from p+p to d+Au, and particularly from peripheral to central d+Au collisions. On the contrary, the away-side peak shows strong differences from peripheral to central d+Au collisions. Peripheral d+Au collisions show back-to-back peak like in p+p, even though it appears to be smaller relative to the near-side peak than in p+p. Central d+Au collisions show instead large broadening of the away-side peak, effectively causing its disappearance. 

Theoretical expectations for azimuthal correlations in interactions between a dilute system (deuteron) and a dense target (Gold) have been explored. Inclusive particle production has been calculated within the CGC framework, using a fixed saturation scale $Q_{S}$ \cite{Albacete}, and then used to compute the di-hadron correlations \cite{cyrille}. Calculations consider valence quarks in the deuteron scattering off low-$x$ gluons in the nucleus with impact parameter $b=0$. A preliminary comparison with data in central d+Au collisions is shown on the right-hand plot of figure \ref{central}. Calculations (in blue) have been superimposed to the data after adding a constant offset to emulate the background from underlying event. CGC calculations  show qualitative consistency with data in their expectations of a strong suppression of the away-side peak in central d+Au collisions. Comparison between same data and CGC calculations using a different approach can be found in this reference \cite{Tuchin}. 

Further systematic studies are being performed for this analysis. Systematic Pythia investigation found p+p data to be consistent with gluon distribution function that include a rapid rise of the gluon density at low $x$. Azimuthal correlations were also studied in embedded Pythia+GSTAR events into minimum bias d+Au data in order to rule out the possibility that additional multiplicity in d+Au events compared to p+p could cause loss of correlation.

\section{Conclusions}
Thanks to a rich d+Au RHIC run in 2008, the Forward Meson Spectrometer is pursuing its primary objective of mapping the boundaries for saturation signatures for back-to-back jet correlations as a function of $\eta$ and $p_{T}$. A new, interesting piece is provided by correlations between two forward neutral pions, which show a strong suppression in the away-side peak in central d+Au collision compared to p+p, qualitatively consistent with CGC expectations.




\section*{References}
\begin{thebibliography}{99}
\bibitem{future}L.C. Bland {\it et al}, \Journal{\EPJC}{43}{427}{2005}.
\bibitem{satu} L. Gribov, E. Levin, and M. Ryskin, \Journal{\PR}{100}{1}{1983}; A.H. Mueller and J.Qiu, \Journal{\NPB}{268}{427}{1986}; L. McLerran and R. Venugopalan, \Journal{\PRD}{49}{3352}{1994}.
\bibitem{GSV} V. Guzey, M. Strickman, and W. Voegelsang, \Journal{\PLB}{603}{173}{2004}.
\bibitem{Qiu}J. Qiu and I. Vitev, \Journal{\PRL}{93}{262301}{2004}.
\bibitem{saturation} A.Dumitru and J. Jalilian-Marian, \Journal{\PRL}{89}{022301}{2002}; E. Iancu, K. Itakura, and D. N. Triantafyllopoulos, \Journal{\NPA}{742}{182}{2004}. 
\bibitem{Albacete} J. L. Albacete, C. Marquet, \Journal{\PLB}{687}{174}{2010}. 
\bibitem{fwd} J. Adams {\it et al}, \Journal{\PRL}{97}{152302}{2006}.  
\bibitem{braidot} E. Braidot, \Journal{\NPA}{830}{603}{2009}.
\bibitem{hijing} X. Wang and M. Gyulassy, \Journal{\PRD}{44}{3501}{1991}.
\bibitem{cyrille} C. Marquet, \Journal{\NPA}{796}{41}{2007}.
\bibitem{Tuchin} K. Tuchin, arXiv:0912.5479v1, (2009).


\end{thebibliography}
\end{document}